\documentclass[%
 reprint,
 amsmath,amssymb,
 prl,
]{revtex4-2}

\usepackage{enumitem}% to have roman letters for enumerate
\usepackage{graphicx}% Include figure files
\usepackage{dcolumn}% Align table columns on decimal point
\usepackage{bm}% bold math
\usepackage{hyperref}% add hypertext capabilities
\usepackage{physics}
\usepackage{array,multirow} % to be able to have more complex tables
\usepackage[dvipsnames]{xcolor} %to be able to add color to text

\newcommand{\minitab}[2][l]{\begin{tabular}{#1}#2\end{tabular}} % To create the table rows split of more rows

\newcommand{\lhm}{\lambda_{\textrm{hm}}}
\newcommand{\lfs}{\lambda_{\textrm{fs}}^{\textrm{eff}}}
\newcommand{\mth}{m_{\textrm{thWDM}}}
\newcommand{\msn}{m_{\textrm{sn}}}
\newcommand{\mhm}{m_{\textrm{hm}}}

\newcommand{\Om}{\Omega_{\textrm{M}}}
\newcommand{\Odm}{\Omega_{\textrm{DM}}}

\newcommand{\AI}{3.81 \cdot 10^8}
\newcommand{\AII}{3.04 \cdot 10^8}

\newcommand{\upperLimitHM}{8.1}
\newcommand{\lowLimitWDMI}{4.6}
\newcommand{\lowLimitWDMII}{4.3}
\newcommand{\lowLimitSnPKI}{11}
\newcommand{\lowLimitSnPKII}{9.8}
\newcommand{\lowLimitSnKTYI}{2.1}
\newcommand{\lowLimitSnKTYII}{1.9}

\newcommand{\lowLimitSnDWI}{34}
\newcommand{\lowLimitSnDWII}{31}
\newcommand{\lowLimitSnNuMSM}{7.0}

\newcommand{\GaHM}{7.0}
\newcommand{\GaWDMI}{9.8}
\newcommand{\GaWDMII}{9.2}
\newcommand{\GaPKI}{26}
\newcommand{\GaPKII}{24}
\newcommand{\GaKTYI}{5.3}
\newcommand{\GaKTYII}{4.9}

\newcommand{\GaDWI}{92}
\newcommand{\GaDWII}{84}
\newcommand{\GaNuMSM}{16}

\newcommand{\LyWDMv}{3.3}
\newcommand{\LyWDMi}{5.3}
\newcommand{\LyHMvI}{8.6}
\newcommand{\LyHMvII}{8.5}
\newcommand{\LyHMiI}{7.9}
\newcommand{\LyHMiII}{7.8}
\newcommand{\LyPKv}{7.1}
\newcommand{\LyPKi}{12}
\newcommand{\LyKTYv}{1.3}
\newcommand{\LyKTYi}{2.5}

\newcommand{\LyDWv}{21}
\newcommand{\LyDWi}{40}
\newcommand{\LyNuMSMvI}{5.0}
\newcommand{\LyNuMSMvII}{5.0}
\newcommand{\LyNuMSMiI}{9.0}
\newcommand{\LyNuMSMiII}{10}

\definecolor{offblue}{rgb}{0.26,0.4,0.74}

\begin{document}

\preprint{APS/123-QED}

\title{Constraints on sterile neutrino models from strong gravitational lensing, Milky Way satellites, and Lyman - \texorpdfstring{$\alpha$}{alpha} forest}% Force line breaks with \\

\author{Ioana A. Zelko$^1$}
\homepage{Corresponding author, ioana.zelko@gmail.com}
\homepage{The code and data for this project can be found at https://github.com/ioanazelko/sterile-neutrinos-constraints.git }
% couldn't get href or url to work here..
\author{Tommaso Treu$^1$}

\author{Kevork N.\ Abazajian$^2$}

\author{Daniel Gilman$^3$}

\author{Andrew J. Benson$^4$}

\author{Simon Birrer$^{5,6}$}

\author{Anna M.~Nierenberg$^7$}

\author{Alexander Kusenko$^{1,8}$}

% \affiliation{%
%  Authors' institution and/or address\\
%  This line break forced with \textbackslash\textbackslash
% }%
\affiliation{[1]
Department of Physics and Astronomy, University of California-Los Angeles,
475 Portola Plaza, Los Angeles, CA 90095
}%

\affiliation{[2] Department of Physics and Astronomy, University of California, Irvine, Irvine, CA 92697, USA}

\affiliation{[3] Department of Astronomy and Astrophysics, University of Toronto, 50 St. George Street, Toronto, ON, M5S 3H4, Canada}

\affiliation{[4] Observatories of the Carnegie Institution for Science, 813 Santa Barbara Street, Pasadena, CA 91101}

\affiliation{[5] Kavli Institute for Particle Astrophysics and Cosmology and Department of Physics, Stanford University, Stanford, CA 94305, USA}
\affiliation{[6] SLAC National Accelerator Laboratory, Menlo Park, CA, 94025}

\affiliation{[7] University of California Merced, Department of Physics 5200 North Lake Rd. Merced, CA 9534}

\affiliation{[8] Kavli IPMU (WPI), UTIAS, The University of Tokyo, Kashiwa, Chiba 277-8583, Japan}

\date{\today}% It is always \today, today,

\begin{abstract}
The nature of dark matter is one of the most important unsolved questions in science. Some dark matter candidates do not have sufficient nongravitational interactions to be probed in laboratory or accelerator experiments.  It is thus important to develop astrophysical probes  which can constrain or lead to a discovery of such candidates. We illustrate this using state-of-the-art measurements of strong gravitationally-lensed quasars to constrain four of the most popular sterile neutrino models, and also report  the constraints for other independent methods that are comparable in procedure. First, we derive effective relations to describe the correspondence between the mass of a thermal relic warm dark matter particle and the mass of sterile neutrinos produced via Higgs decay and GUT-scale scenarios, in terms of large-scale structure and galaxy formation astrophysical effects. Second, we show that sterile neutrinos produced through the Higgs decay mechanism are allowed only for mass $>\GaPKI$ keV,  and GUT-scale scenario $>\GaKTYI$ keV. Third, we show that the single sterile neutrino model produced through active neutrino oscillations is allowed for mass $>\GaDWI$ keV, and the 3 sterile neutrino minimal standard model ($\nu$MSM) for mass $>\GaNuMSM$ keV. These are the most stringent experimental limits on
these models.
%the Higgs decay, GUT-scale scenarios, and the 3 sterile neutrino minimal standard model, and also provide a completely independent limit on the single neutrino active neutrino osicallations production model.
\end{abstract}

\maketitle

%\tableofcontents

\section{Introduction}\label{sec:introduction}

%Multiple observations imply that the mass content of the univer-se is dominated by an unknown type of matter \citep{PlanckCollaboration2020}, which contributes $\sim$25\% of the total energy. This matter is not made of ordinary atoms; it has no significant electromagnetic interaction,  and it is thus called ``dark matter'' (DM). The nature of DM is one of the most important questions in modern science, with critical implications spanning from particle physics to astrophysics and cosmology.

The nature of dark matter (DM) is one of the most important questions in modern physics, with implications spanning from particle physics to astrophysics and cosmology. This unknown particle contributes 25\% of the total energy of the universe \citep{PlanckCollaboration2020}, but is not made of ordinary matter and has no electromagnetic interaction.

%Dark matter models have a great impact on the evolution of the universe, and the structure in it. It is currently stipulated (citations) that dark matter drives the formation of galaxies;  as as the universe evolves, the initial inhomogeneities in the matter density field lead to dark matter clumping together forming dark matter halos, which in turn impact the clumping of baryonic matter which leads to the formation of galaxies.

Many DM models have been proposed. A number of candidates fall into the class of cold dark matter (CDM)~\cite{Blumenthal1984}, made of collisionless particles considered ``cold'' due to their small velocity dispersion relative to the speed of light. This model is extremely successful on supergalactic scales but there are open challenges at subgalactic scales \citep{Bullock2017}. %For example,
CDM predicts more satellites than are observed around galaxies of Milky Way (MW) mass, ``cuspy'' dark matter density profiles in contrast to the flatter cores observed in dwarf galaxies and clusters, and predicts that subhalos hosting the largest MW satellites are either underdense or too small. It is still unclear whether these challenges can be solved by a better understanding of baryonic processes, or whether alternative dark matter models are needed \citep[e.g.][and references therein]{Weinberg2015}.

%\textcolor{blue}{add the references for the adjustments made to CDM that fix these?}

Plenty of DM models have been proposed to eliminate these small-scale tensions between observations and CDM \citep{Seigar2015}. 
DM particles which are generated with higher velocity dispersions erase fluctuations in the matter power spectrum at scales smaller than a characteristic `free-streaming length', suppressing structures below this scale. 
So-called `hot' dark matter candidates such as standard neutrinos 
%with relativistic speeds would have free-streaming lengths that would erase structure on galactic scales and 
are ruled out by observations \cite{Gerhard1992, Primack1995,
Menci2017,Hsueh2020}, as the main DM component. However, a broad range of ``warm'' DM (WDM) with smaller but non-negligible free streaming lengths are viable. One popular class of WDM models are sterile neutrinos (SNs).

%WDM models (WDM) are an intermediate class of models with velocity dispersion between these limits. 

%One such class of models are the sterile neutrinos.

Sterile neutrinos~\cite{Kusenko2009,Abazajian2012,Abazajian2017,Adhikari2017,Boyarsky2019} are %a hypothetical class of particles that can be very promising models for dark matter. They are neutrino
particles with right-handed chirality, no charge, and no color charge, and therefore do not interact  with standard model particles except via mixing with neutrinos, or via some non-standard-model interactions.
%, which so far has been known to couple only to particles with negative spins. 
They were first introduced for the purpose of explaining the masses of active (left-handed) neutrinos, and can have masses in the range from eV to the Planck scale.
In the early universe they can decouple from the plasma before electron-positron annihilation, when they are still relativistic \citep{Malaney1991, Hannestad2012,Abazajian2017, Gariazzo2019, Alonso-Alvarez2022}.

%Nearly three decades ago, 
The exact production mechanism for a given SN model determines the clustering of dark matter \citep{Kusenko2006,Petraki2008,Petraki:2008ef,Abazajian2019}. \citet{Dodelson1994} (hereafter the DW) model   %first established the SN as a well-motivated WDM candidate:
%dark matter particle candidate that would behave on cosmological scales as warm dark matter. 
adds a single SN with coupling to active neutrinos. DW SNs are produced in neutrino oscillations at temperatures below a few GeV \cite{Abazajian2006}. %SNs were shown to be a good warm dark matter candidates first through the Dodelson-Widrow (hereafter DW) model \cite{Dodelson1994}, which relies on a single particle being added to the Standard Model, the SN, coupling to the active  neutrinos. They are produced through  neutrino oscillations, at temperatures below GeV, and thus are in the WDM category.
%add constraints that already exist on this model

Alternatively, there may be multiple SN species produced in neutrino oscillations. In the seesaw theory of neutrino masses, one needs at least two right-handed states~\cite{Frampton:2002qc}, but more right-handed states are allowed. In the presence of a sizeable lepton asymmetry, the active to SNs conversions select a lower-momentum part of the thermal distribution, leading to somewhat colder dark matter
%, as was pointed out by 
~\citep[Shi and Fuller (SF),][]{Shi1999}.  Thus, the popular neutrino minimal standard model \citep[][hereafter $\nu$MSM or SF]{Shi1999,Asaka2005} postulates three right-handed neutrinos with masses below the electroweak scale. 

%The number of SNs is not subject to anomaly cancellation or other constraints, and more than one SN is possible.  In the seesaw theory of neutrino masses, one needs at least two right-handed states~\cite{Frampton:2002qc}, but more right-handed states are allowed.  It is an open question how many of these states have masses in the keV range, where they can serve as WDM. 

%%% this figure is called here to help with the placement on the next page
% \begin{figure*}[t!]
%     \hspace{-7mm}
% 	\includegraphics[width=\linewidth]{joint_posterior.pdf}
% 	\caption{Joint distribution posterior from analysis of all the quasars as performed in \cite{Gilman2020}. [TT: I don't think we need this figure unless we explain what is on the x-axis and that seems too much detail] \label{fig:joint_posterior}}
% \end{figure*}

SNs could also be produced through mechanisms other than active neutrino oscillations, including "freeze-in" production from decays of the inflaton~\cite{Shaposhnikov:2006xi} or an SU(2)$\times$U(1) singlet Higgs boson ~\citep[PK,][]{Kusenko2006,Petraki2008,Petraki:2008ef,Abazajian2019}. 
Most of the SNs production from oscillations (DW or SF) takes place at temperature $\sim 0.1$~GeV.  In contrast, Higgs boson decays can produce a population of SNs at a temperature $\sim$100~GeV. Subsequent cooling and entropy production dilutes and redshifts this population, making the resulting DM colder than the WDM produced by DW or SF. This model, which produces particles in the 1-10 keV range, has been shown  to produce the correct dark matter abundance, resulting in the so-called ``keV Miracle Model'' (hereafter PK).

Another possible production mechanism for SNs is the split seesaw mechanism \citep[hereafter KTY]{Kusenko2010}. %part of the production mechanism at the grand unified theory scale (GUT). 
The model predicts two large and  one small Majorana masses due to a natural separation of scales. The large Majorana masses allow for thermal
leptogenesis, while the keV mass produces a DM candidate. The model can be embedded into an SO(10) Grand Unified Theory (GUT), or some other theory containing a gauge U(1)$_{B-L}$ symmetry. The resulting DM is colder than the models described above, due to dilution and redshifting of SNs as the plasma cools from the GUT-scale production temperature. 

%SN WDM models offer potential solutions to the dark matter problem. However, due to their energy range and production mechanisms, their production rate in traditional particle physics experiments would be too small for detection \textcolor{blue}{need citation, I do not know it}.

%To solve this problem, different methods that rely on cosmological observations can be used. 

The 4 SN models described above could explain the unidentified 3.5 keV X-ray line found in observations of galaxies and clusters~\cite{Bulbul:2014sua, Boyarsky:2014jta}, even if they do not account for 100\% of DM.  Furthermore, keV SNs can have a dramatic effect on supernovae, e.g., explaining the pulsar kick velocities in excess of 1000~km/s which so far has evaded other explanations~\cite{Kusenko:1997sp,Fuller:2003gy}. 

We use state-of-the-art measurements of gravitationally lensed quasars \citep{Treu2010,Gilman2020}, MW satellites \citep{Schneider2016, Cherry2017, DESCollaboration2021, Nadler2021}, and Ly$\alpha$ \citep{Viel2013, Irsic2017} , to constrain the 4 popular SN models described above. 
%In practice, we build on the work by \cite{Gilman2020} who analyzed and modeled the measurements obtained by \citep{Nierenberg2020,Nierenberg2017,Nierenberg2014} to infer the the number of dark matter halos as a function of halo mass (hereafter the subhalo mass function). They used their inference to constrain a standard reference warm dark matter model composed of thermal relic. We also use the results of \cite{Nadler2021}, who combined this strong lensing data sets with priors based on astrophysical constraints from the MW satellite counts, and obtained even tighter limits.  Theirs are the most stringent astrophysical limits to date on thermal relic mass, comparable and completely independent from those obtained by studying the Ly$\alpha$ forest. 
We take their limits in terms of the thermal relic WDM, and compute the equivalent limits for the four SN models described above, 
%We take their limits in terms of a traditional reference WDM model composed of a thermal relic, and  compute the equivalent SN models, 
from the point of view of cosmological structure formation and the halo mass function.  As we will show, our analysis provides the most stringent limits to date on these four SN models.

\section{Thermal Relic Warm Dark Matter Constraints from Strong Gravitational Lensing}\label{sec:strong_lensing_data}

% \begin{figure}[t]
% \includegraphics{m_hm_posterior_log.pdf}
% \caption{\label{fig:m_hm_posterior_log} Marginalized posterior for the $\log_{10}(m_{\textrm{hm}})$, obtained by \cite{Gilman2020} from the analysis of eight quadruply imaged quasars. \textbf{The vertical dashed line marks the 95\% upper boundary interval.}}
% \end{figure}

% \begin{figure}[t]
% \includegraphics{m_thWDM_p_norm.pdf}% Here is how to import EPS art
% \vspace{-2mm}
% \caption{\label{fig:m_WDM_p_norm} Marginalized posterior of the masses of the thermal relic WDM particle.
% %, obtained from the marginalized posterior of the half-mode mass shown in Fig. \ref{fig:m_hm_posterior_log} using Eq. \ref{eq:p_wdm}. 
% The vertical \textbf{dashed} line marks the 95\% lower boundary interval, corresponding to \lowLimitWDMI  ~and \lowLimitWDMII keV, which represent the case where the assumption for the average background density of the universe includes (Case I), or does not include (Case II), baryonic matter in addition to dark matter (see Appendix).}
% \vspace{-5mm}
% \end{figure}

\begin{figure}[t]
\includegraphics[width=\columnwidth]{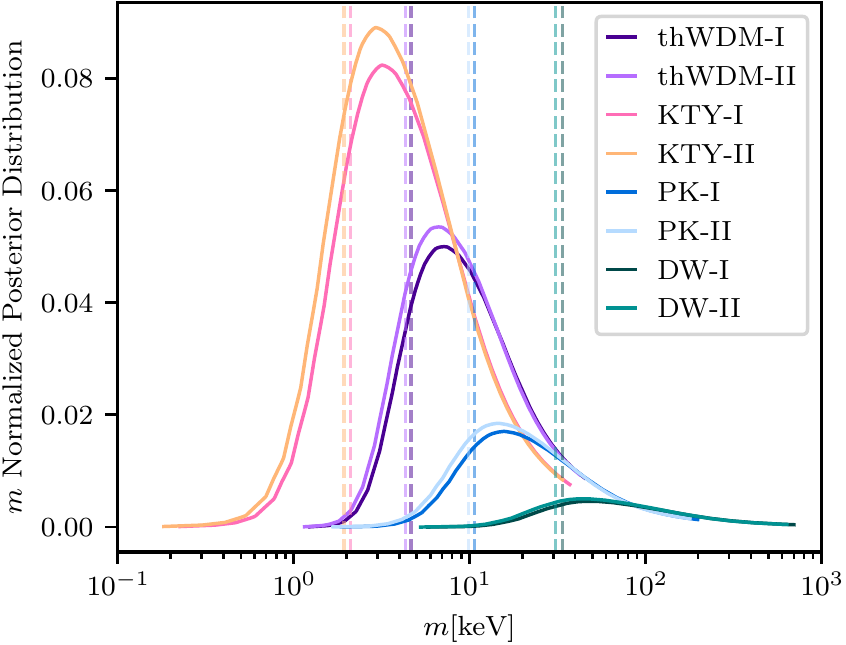}% Here is how to import EPS art
\vspace{-5mm}
\caption{\label{fig:m_sn_p_norm} Posterior probability distribution function $p(m)$ of the mass of thermal relic WDM and the various kinds of SN: the GUT-scale scenario (KTY), the `keV miracle model' Higgs production mechanism (PK), and single particle neutrino oscillation production mechanism (DW). The posteriors do not go to 0 on the right limit, so we cannot impose upper constraints on the particle masses; however, since they do go to 0 on the small limit, we can derive a lower limit.
The vertical dashed line marks the 95\% lower boundary interval, corresponding to  $\lowLimitWDMI$/$\lowLimitSnKTYI$/$\lowLimitSnPKI$/$\lowLimitSnDWI$ 
keV for thWDM/KTY/PK/DW. The limits for the $\nu$MSM model depend on lepton asymmetry, and are discussed in the text. The case where the assumption for the average background density of the universe includes (Case I), or does not include (Case II), baryonic matter in addition to dark matter (see Appendix) is shown. 
}
\vspace{-5mm}
\end{figure}
%Using the thermal relic WDM transfer function equation (Eq. \ref{eq:wdm_transfer_f}), 

%Galactic halos can be dark, devoid of light-emitting/absorbing matter. 
Strong gravitational lensing depends only upon gravity and is thus sensitive to the abundance of halos irrespective of their ability to emit or absorb light.% making it a very powerful probe of the universe on subgalactic scales.  
It can thus determine the halo mass function directly, avoiding uncertainties related to the physics of star formation in low mass galaxies that affect traditional methods.
%based on counting photons.
%and can separate baryonic physics from the nature of dark matter. It is also able to improve limits on constrains for dark matter models. 

\begin{figure}[t!]
	\includegraphics{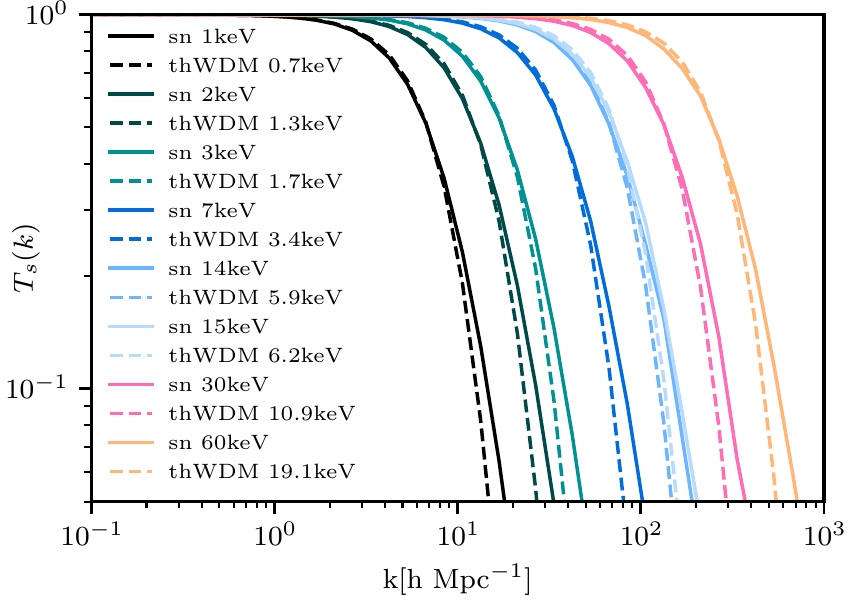}
	\vspace{-8mm}
 	\caption{SN transfer functions belonging to the Higgs decay (PK) model proposed by \cite{Petraki2008}, shown by the continuous lines. The dashed lines show the corresponding thermal relic WDM transfer functions. 
 	 %obtained by fitting Eq. \ref{eq:wdm_transfer_f} from \cite{Bode2001} to each sn transfer function.
    As shown in \cite{Abazajian2019}, the two sets of functions are very similar to each other, thus allowing the possibility to create a mapping between the masses of thermal relic WDM particles and those of SNs. \label{fig:transfer_functions_WDM_fit}}
    \vspace{-5mm}
\end{figure}

\cite{Gilman2020} used eight quadruply-imaged quasars systems to constrain the amplitude of the halo mass function and the free-streaming length of dark matter. For each system, many realizations of dark matter structure are drawn from analytic dark matter halo mass functions flexible enough to describe mass functions produced by a broad range of thermal relic dark matter masses.

The predicted flux ratios for each realization are compared with the observed flux ratios to estimate the likelihood using Approximate Bayesian Computing. 
%Dark matter models which are more likely to produce the observed image fluxes are thus statistically more probable.
The likelihoods from each system are multiplied together to infer the parameters common to all systems. A key parameter 
%in the halo mass function inference 
is the `half-model mass' $\mhm$ which is the mass at which there are half as many halos as there would have been in the case of CDM. 

 We marginalize over all the other parameters 
 %in their analysis
 %, and assume a prior linear in $\log10{\mhm}$, 
 to obtain the posterior distribution for $\mhm$. % shown in Fig. \ref{fig:m_hm_posterior_log}. 
%m_hm = 10**log10_m_hm
%We convert between the logarithmic and linear posterior distributions.
As always in Bayesian statistics, the posterior depends on the choice of priors. Since we do not know the order of magnitude of the SN mass we adopt a uniform prior in $\log10{(\mhm)}$, within the range $10^{4.8}-10^{10} M_{\odot}$ (reported in units of Sun mass following existing literature on the halo mass function).
% \begin{equation}
%     m_{\textrm{hm}} = 10^{\log_{10}{m_{\textrm{hm}}}}
% \end{equation}

% \begin{equation}\label{eq:p_log_to_linear}
%     p( m_{\textrm{hm}})  =  \frac{p(\log_{10}{m_{\textrm{hm}}})}{m_{\textrm{hm}} \ln{10}}
% \end{equation}

The constraints on $\mhm$ can be tied into constraints on the mass of the DM particle, given a model. A traditional reference model is the thermal relic WDM. It does not refer to a physical particle in particular, but serves as a standard tie-in model for the properties of WDM models with thermal relics. \cite{Bode2001}, \cite{Schneider2012} derived a one-to-one mapping between the half-mode mass and the mass of the thermal relic WDM. The general form of the conversion is 

%we show the derivation of the two cases for the relation, which depend on our assumptions.

\begin{equation}\label{eq:m_WDM_to_m_hm}
    m_{\textrm{thWDM}}  = 3.3 \left[\left(\frac{m_{\textrm{hm}}[M_{\odot}]}{A}\right)^{-1/3.33}\right]\textrm{keV} ,
\end{equation}
%
%
%p_WDM_d = 3E8 p_m_hm/(3.3/3.33*(m_hm/3E8)**(-1/3.33-1))
% \begin{equation}\label{eq:p_wdm}
%     p(m_{\textrm{thWDM}}) = p(m_{\textrm{hm}} ) \frac{  A }{ 3.3/3.33  } \left(\frac{m_{\textrm{hm}}[M_{\odot}]}{A}\right)^{4.33/3.33},
% \end{equation}
% %
where $A$ has two possible values (Case I and II), depending on assumptions about the background density of the universe, as detailed in the Appendix. Using Eq. \ref{eq:m_WDM_to_m_hm}, 
%and \ref{eq:p_wdm}, 
we obtain the posterior %on $\mth$ can be obtained (
%shown in Fig.~\ref{fig:m_WDM_p_norm}.
shown in Fig.~\ref{fig:m_sn_p_norm}.

\section{Relation between thermal relic WDM and sterile neutrino transfer functions}\label{sec:transfer_functions}

Our goal is to use the 
%thermal relic WDM and half-mode mass 
constraints discussed as an illustration in \S \ref{sec:strong_lensing_data}, and those obtained by \citep{Nadler2021} and \cite{Viel2013, Irsic2017} to derive constraints on the four SN dark matter candidates discussed in \S \ref{sec:introduction}: the GUT-scale scenario (KTY), the `keV miracle model' Higgs production mechanism (PK), the single particle neutrino oscillation production mechanism (DW), and the Shi-Fuller mechanism within the neutrino minimal standard model ($\nu$MSM).

The key quantity
%ingredient for this conversion 
is the transfer function, $T$, which describes the effect of free-streaming on matter distribution. 
 %The transfer function $T$ is the key link between dark matter properties and growth of structure. 
Given the power spectrum of initial density fluctuations $P_i$, $T$ describes its evolution as a function of scale $k$ and cosmic time, with respect to a standard CDM model:  $T_s(k)\equiv \sqrt{\frac{P_{\textrm{sterile,i}}(k)}{P_{\textrm{CDM,i}}(k)}}$.

% %of the power spectrum of density fluctuations on scale $k$ 
% \begin{equation}
%     P(k,t) = P_i(k) T^2(k)D^2(t),
% \end{equation}
% %
% %
% \begin{equation}
%     T_s(k)\equiv \sqrt{\frac{P_{\textrm{sterile,i}}(k)}{P_{\textrm{CDM,i}}(k)}},
% \end{equation}
% where $D(t)$ encodes the growth with time.

% Our goal is to use relationships between the transfer functions to convert the strong lensing constraints from \cite{Gilman2020} into constraints on four SN models discussed in Section \ref{sec:introduction}: t

%%%%%%%%%%%%%%%%%%%%%%%%%%%%%%%%%%%%%%%%%%%%
%%%% Input the table of the fit coefficients
%%%%%%%%%%%%%%%%%%%%%%%%%%%%%%%%%%%%%%%%%%%%
\begin{table}[t]
\renewcommand{\arraystretch}{1.3}% for the vertical padding
\setlength{\tabcolsep}{3pt}
\centering
\begin{tabular}{lllll}
\hline
\hline
 & deg ~  & $a_0$~ & $a_1$~ & $a_2$~ \\ 
\hline
\multirow{2}{*}{PK} &   1st &-3.26e+00&3.21e+00\\ 
& 2nd &-1.06e+00&2.31e+00&4.66e-02\\ 
\hline
\multirow{2}{*}{KTY} &   1st &-1.10e+00&6.87e-01\\ 
& 2nd &-4.17e-01&5.11e-01&7.15e-03\\ 
\hline
power & law ~  & $a$~ & $b$~ \\ 
\hline
PK& &1.56e+00&1.24e+00\\ 
\hline
KTY& &3.14e-01&1.24e+00\\ 
\hline
\end{tabular}
\caption{Coefficients for the fits for the relation between the $\msn$ and $\mth$, for the cases of the Higgs production mechanism (PK) and GUT scale (KTY). The power law fit, as well as the first 2 degree polynomials are shown. \label{table:fit_coeffs}}
\end{table}

Transfer functions for SNs (Fig. \ref{fig:transfer_functions_WDM_fit}) for the Higgs production mechanism and the GUT-scale scenarios have been previously obtained by \cite{Abazajian2019}. We calculate the transfer functions of the several models by using the momentum-space distribution functions as tables that are provided to CLASS. Including these with the proper effective temperature of the dark matter models allows for accurate calculation of the transfer functions relative to CDM. The ones given here supersede the published ones by adopting more up to date cosmological parameters. We also corrected a mismatch between expected and provided dilution factors, using the non-CDM features of CLASS 
%to calculate the transfer functions
\cite{Blas2011}. For the initial conditions, we fix the cosmology to the mean results from \cite{PlanckCollaboration2020} and the CMB temperature from \cite{Fixsen2009}.%: $H_0=67.36$, $\Omega_{\textrm{Matter}}=0.31530$, $\Omega_{\Lambda}=0.68470$, $\Omega_{\textrm{Baryon}}=0.04930$, $\Omega_{\nu} = 0.00142$, $T_{\textrm{CMB}}=2.72548$K.
%these functions indicate a one-to-one correspondence between the masses of thermal $\mth$ particles and SNs for which the effect  on the matter distribution and thus the transfer function for both models are similar.

%{Mapping unto thermal DM}
%The advantage of having such ha mapping between SNs and thermal relic warm dark matter is that when we design experiments, we don't want to redo all the experiments that have already been done.
%If we have the mapping from thermal relic to a model, it is very simple to impose constraints on that model.

\begin{figure}[t]
\includegraphics{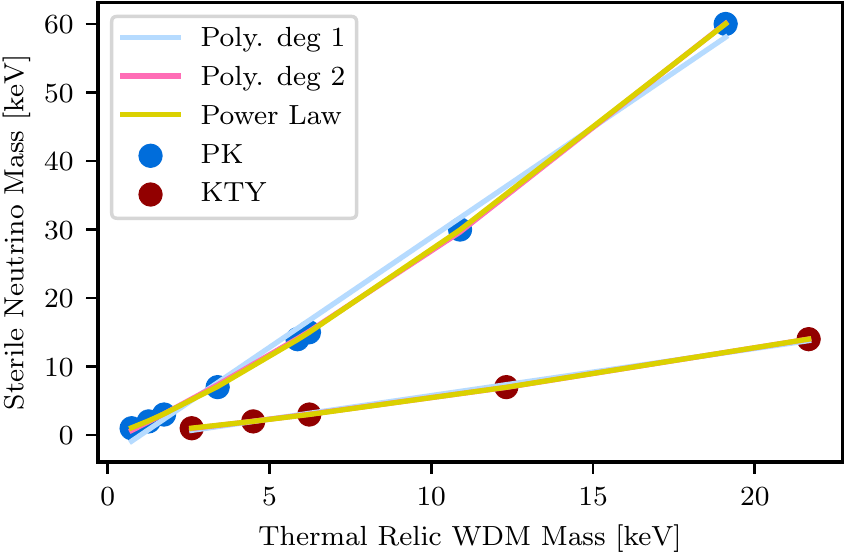}% Here is how to import EPS art
\vspace{-4mm}
\caption{\label{fig:sn_WDM_relation} The masses of the thermal relic WDM particles and SN particles corresponding to the PK transfer functions shown in Fig.~\ref{fig:transfer_functions_WDM_fit}, as well as those for the KTY model, are shown as scattered points. Polynomial and power-law fits %described by Eqs.~\ref{eq:m_sn_vs_m_WDM} and \ref{eq:m_sn_vs_m_WDM_power} 
are shown as solid lines. These relations allow us to map constraints on the mass of thermal relic WDM to the corresponding mass of a SN, for the Higgs production mechanism (PK), and the GUT-scale scenario (KTY). }
\vspace{-5mm}
\end{figure}

The thermal relic WDM transfer functions for a given $\mth$ can be obtained from the analytical form presented in equation A9 of \cite{Bode2001} (with more recent numbers from \cite{Viel2005}), and it is given by
%The transfer function for thermal  relic WDM is given by 

\begin{equation}\label{eq:wdm_transfer_f}
    T_{\textrm{thWDM}}(k) = \left[1+ (\alpha k)^{2\mu} \right]^{-5/\mu}
\end{equation}
with $\alpha, \mu$ given in the Appendix.

We tested two methods to find the relation between SN and thermal relic WDM transfer functions: first, we fit the SN transfer functions using the mass of the thermal relic WDM as a free parameter (Eq. \ref{eq:wdm_transfer_f}); second, we match the 'half-mode' wavenumber where the transfer functions decrease to 0.5.
%For a given thermal relic WDM particle mass and transfer function, 
The half-mode $k_{\textrm{hm}}$ and the mass of the particle $\mth$ can be related analytically from Eq. \ref{eq:wdm_transfer_f}, to obtain:

\begin{equation}
\begin{split}
    \mth = & k_{\textrm{hm}}^{\frac{1}{1.11}} \left(0.049 \left(2^{\frac{\mu}{5}}-1\right)^{-\frac{1}{2\mu}} \times \right.\\
    & \left. \left(\frac{\Omega_X}{0.25}\right)^{0.11} \left(\frac{h}{0.7}\right)^{1.22} \right)^{\frac{1}{1.11}}.
\end{split}
\end{equation}
In the first method, the fits might depend too much on how the numerical $T(k)$ for the SNs was sampled, and the errors that arise at high $k$. The 1/2 mode matching gets rid of those issues, so we use this one for the rest of the paper. The difference in the end results of the two methods is of order $1\%$.
We obtain the fits shown in Fig. \ref{fig:transfer_functions_WDM_fit}. For root-mean-square goodness of fit estimator, the PK model KTY model were comparable.

The relation between the mass of the SN which generates an equivalent transfer function to a thermal relic particle with a given mass can be approximated by a polynomial of degree $deg$, $m_{\textrm{sn}} = f(\mth) = \sum_{0}^{deg} a_i\cdot\mth^i$,
% \begin{equation}\label{eq:m_sn_vs_m_WDM}
%     m_{\textrm{sn}} = f(\mth) = \sum_{0}^{deg} a_i\cdot\mth^i,
% \end{equation}
%where $deg$ is the degree of the polynomial being fit, 
or by a power law $ m_{\textrm{sn}} = a\cdot\mth^b$.
% \begin{equation}\label{eq:m_sn_vs_m_WDM_power}
%     m_{\textrm{sn}} = a\cdot\mth^b
% \end{equation}

Results of the polynomials fits  of orders 1, 2 (higher orders resulted in over fitting), as well as the power law, can be seen in Table \ref{table:fit_coeffs}.
Using these coefficients, a relation can also be derived between $\msn$ and $\mhm$, seen below in the case of the power law fit:
\begin{equation}\label{eq:m_sn_vs_m_hm_power}
    \msn = a \cdot 3.3^b \left(\frac{\mhm}{A}\right)^{-b/3.33}
\end{equation}
where A takes the values described in the Appendix.
These relations are calibrated on the $\mth$ mass interval of (0.75, 22) keV, and then extended to 60 keV when applied to data. 
Fig. \ref{fig:sn_WDM_relation} shows the results of these fits.
%For the rest of the paper, 
We use the results of the second order polynomial 
%since it provides the better fit, 
but we note that using a linear approximation would change the inferred bounds on SN mass by only 11\%.

% Remarkably, the transfer functions for thermal particles and SNs are very similar, and one can approximate the other with a simple mapping of the relevant mass. 
This mapping allows one to convert any results obtained for thermal relic WDM to SN in post-processing, without redoing the experiment or the analysis.

For SNs produced through the oscillation mechanism for the Dodelson-Widrow model, the relation between $\msn$ and $\mth$ is taken from \cite{Viel2005}. 

For the $\nu$MSM model, \cite{Vegetti2018} derives the model connections to the half-mode mass $\mhm$, and we use the $\mhm$ posterior 
%in Fig. \ref{fig:m_hm_posterior_log} 
to constrain them.
%\begin{equation}\label{eq:posterior_sm_wDM}
%    p(m_\textrm{sn}) = \frac{p(m_\textrm{thWDM})}{|\dv{\msn}{\mth}|} = %\frac{p(m_\textrm{thWDM})}{|\sum_{1}^{deg }a_i i \mth^{i-1} |}
%\end{equation}

\section{Mapping Thermal Relic Warm Dark Matter Constraints onto sterile neutrinos}\label{sec:constraints_on_sterile_neutrinos}

We obtain the most stringent
experimental limits on four SN models: PK, KTY, $\nu$MSM, and DW.

To recap, our starting point are the following 95\% confidence limits on thermal relic WDM. The lensing-only analysis described in \S \ref{sec:strong_lensing_data} gives $m_\mathrm{thWDM} > \lowLimitWDMI \mathrm{keV}$ \cite{Gilman2020}. Combination with satellite counts extends it to $m_\mathrm{thWDM} > 9.7\, \mathrm{keV}$ \cite{Nadler2021}. Independent work on the Lyman-$\alpha$ forest yields 3.3 keV \citep{Viel2013} and 5.3 keV  \cite{Irsic2017}, the latter using additional assumptions for the relevant thermodynamics. %These models are all assuming standard cosmology

For the PK and KTY models, the relations derived in \S~\ref{sec:transfer_functions} can now be used together with the ones for $\nu$MSM and DW to translate limits from thermal relic WDM into limits for the SNs masses (Fig. \ref{fig:m_sn_p_norm}).  The 95\% limits for the four models 
%under consideration 
are given in Tab.~\ref{table:limits}.

%These become our limits of integration, and thus the priors. 

We note that the posterior shown in Fig.~\ref{fig:m_sn_p_norm}  vanishes at the lower bound but not on the upper bound, as expected because the warmest models are ruled out by a number of observations \citep{Menci2017, Hsueh2020, Nadler2021}. This puts a lot of weight on the choice of priors, which can influence the limits reported. We thus convert the posteriors to lower limits, although of course the full posterior is more informative. Likelihood ratios can be obtained from Fig.~\ref{fig:m_sn_p_norm}.

For the $\nu$MSM model, we use the posterior on $\mhm$ 
%(Fig. \ref{fig:m_hm_posterior_log}) 
to eliminate the model space as shown in Fig. 2 of \cite{Vegetti2018} , which presents the expected half-mode mass as a function of lepton asymmetry (L6) for different neutrino masses. Our upper limit on $\log_{10}( \mhm [M_{\odot}])$ from strong lensing alone is $\upperLimitHM$. This rules out masses under $\lowLimitSnNuMSM$ keV for all lepton asymmetries. For higher masses, only limited ranges of lepton asymmetries are allowed: 7keV: L6 $\in$ (6.8, 7.6), 9keV: L6 $\in$ (5.2,7.8), 11keV:L6 $\in$ (4.3, 7.6), 14keV: L6 $\in$ (1.7, 7.9), 16 keV: L6 $\in$ (1.6, 11.5). After incorporating the MW satellite counts constraints,  $\log_{10}( \mhm [M_{\odot}])>7.0$, which corresponds to $m_{\nu MSM}>$\GaNuMSM keV. These limits are improved compared to existing work (\cite{Dekker2021}). These results are contingent on the assumption that DM is made from a single component, the SN model of choice. However, DM could be a mixture of different components, such as the “mixed cold+warm DM model”\citep{Boyarsky2009,Baur2017}. Adding a parameter to control the abundance ratio could make the constraints weaker.

Future work would aim to combine the limits from strong lensing, galaxy counts, and Lyman-$\alpha$ forest in a joint analysis. Work has already been done in this direction (\cite{Enzi2021}), however their dataset obtained less stringent limits than the strong lensing combined with galaxy counts obtained by \cite{Nadler2021}. Future analysis combining the data sets used in Table \ref{table:limits} will be useful. In addition, work exploring analytical connections in the non-linear regime (as was done at the level of the transfer function by \cite{Murgia2017}) at the sub-halo level would be useful.
\begin{table}[t]
%\footnotesize
\scriptsize
\setlength{\tabcolsep}{1pt}
\renewcommand{\arraystretch}{1.4}% for the vertical padding
\centering
\begin{tabular}{|l|c|c|c|c|}
\hline
\hline
 & \multirow{3}{*}{\minitab[c]{Strong \\ Lensing}}  & Strong Lensing   & \multirow{3}{*}{Lyman-$\alpha$ } & Lyman-$\alpha$ \\ 
 & & \& & & \& \\
 &  &  Galaxy Counts  & & Thermo.\\
%%%%%%%%%%%%%%%%%%%%%%%
\hline
PK [keV] & I: \lowLimitSnPKI,  II: \lowLimitSnPKII &I: \GaPKI ,II: \GaPKII   & \LyPKv  &\LyPKi\\
%%%%%%%%%%%%%%%%%%%%%%%
\hline
KTY [keV] & I: \lowLimitSnKTYI, II: \lowLimitSnKTYII  &I: \GaKTYI,II: \GaKTYII &\LyKTYv & \LyKTYi \\
%%%%%%%%%%%%%%%%%%%%%%%%%
\hline
$\nu$MSM [keV]  &\lowLimitSnNuMSM  &\GaNuMSM & I: \LyNuMSMvI, II: \LyNuMSMvII    &I:  \LyNuMSMiI, II:  \LyNuMSMiII\\
\hline
DW [keV] &I: \lowLimitSnDWI, II: \lowLimitSnDWII &I: \GaDWI, II: \GaDWII &\LyDWv & \LyDWi\\

\hline
\hline

$\log_{10}{(\textrm{hm}[M_{\odot}])}$  &\upperLimitHM &\GaHM & I: \LyHMvI, II: \LyHMvII   &I: \LyHMiI, II: \LyHMiII\\
\hline
thWDM [keV] &I: \lowLimitWDMI,  II: \lowLimitWDMII &I: \GaWDMI, II: \GaWDMII & \LyWDMv& \LyWDMi\\

\hline
\hline

\end{tabular}
\caption{
%We present the 
95$\%$ lower limits for four sterile neutrino models: Higgs production mechanism (PK), GUT scale scenario (KTY), 3 sterile neutrino minimal standard model ($\nu$MSM), single model sterile neutrino produced through active
neutrino oscillations (DW). The limits are derived from 4 datasets: gravitational strong lensing \citep{Gilman2020}, strong lensing combined with Milky Way galaxy counts \citep{Nadler2021}, Lyman-$\alpha$ forest \citep{Viel2013}, and Lyman-$\alpha$ forest combined with thermodynamic assumptions \citep{Irsic2017}. The case I and case II labels correspond to different assumption cases about the average background density of the universe, as described in the Appendix. \textbf{In the last two rows of the table, we also presents the limits in terms of the half-mode mass, and the thermal relic WDM mass.} \label{table:limits}}
\vspace{-7mm}
\end{table}

\section{Conclusion}\label{sec:conclusion}

We used flux ratios of strong gravitationally-lensed quasars, MW satellites, and Lyman-$\alpha$ forest to constrain four of the most popular SN models. 

First, we derive effective relations to describe the correspondence between the
mass of a thermal relic WDM particle and the mass of SNs produced via Higgs decay and GUT-scale scenarios, in terms of astrophysical effects. We take advantage of the similarity between the transfer functions of the SNs mechanism presented by \cite{Abazajian2019}, to that of thermal relic WDM.
 
We note that our derived equivalence relations are of general importance, and can be used to put limits on SN models for any thermal relic WDM measurement, not just the ones we present here.

%Second, we apply the relationships we derived between $\msn$ and $\mth$ to the gravitational strong lensing constraints of \cite{Gilman2020}. We show that sterile
%neutrinos produced through the Higgs decay mechanism are constrained by the lensed quasars data %to have  $m>\GaPKI$ keV,  and GUT-scale scenario $m>\GaKTYI$ keV. 
%With this, we have ruled out part of the parameter space  for sn  generated through the Higgs mechanism and GUT-scale scenario proposed by \cite{Petraki2008}, \cite{Kusenko2010},\cite{Abazajian2019}.

%Third, we show that the 3 sn minimal standard model ($\nu$MSM) is constrained to have mass $>\GaNuMSM$ keV. Finally, we show that the single model sn produced through active neutrino oscillations is constrained to $m>\GaDWI$. 

The limits on the PK, KTY, $\nu$MSM and DW models summarized in Table~\ref{table:limits} are the most stringent experimental limits on these four models. We note that the limits from lensing and MW satellites are independent of and agree with those from the Ly$\alpha$ forest. We have effectively ruled out part of the parameter space for SN generated through these 4 models.

\begin{acknowledgments}

We acknowledge helpful conversations with Graciela Gelmini, Jiamin Hou, Doug Finkbeiner, Ethan Nadler, Joshua Speagle, and Xiaolong Du.

IZ and TT acknowledge support by the National Science Foundation grant NSF-1836016, %``Astrophysics enabled by Keck All Sky Precision Adaptive Optics"
by the Gordon and Betty Moore Foundation Grant 8548 
%``Cosmology via Strongly lensed quasars with KAPA'', 
and by NASA grant HST-GO-15177. Part of the data used in this paper were obtained as part of HST-GO-15177.
SB is supported by the National Science Foundation through NSF AST-1716527. K.N.A.\ is supported in part by NSF Theoretical Physics Grant PHY-1915005. 
A.K. was supported by the U.S. Department of Energy (DOE) grant No. DE-SC0009937 and by Japan Society for the Promotion of Science (JSPS) KAKENHI grant No.
JP20H05853, as well as by World Premier International Research Center Initiative (WPI), MEXT, Japan.   This work was supported in part by the UC Southern California Hub, with funding from the UC National Laboratories division of the University of California Office of the President.
This research made use of the NASA Astrophysics Data System's Bibliographic Services (ADS),  the color blindness palette by Martin Krzywinski and Jonathan Corum\footnote{\url{http://mkweb.bcgsc.ca/biovis2012/color-blindness-palette.png}}, the Color Vision Deficiency PDF Viewer by Marie Chatfield \footnote{\url{https://mariechatfield.com/simple-pdf-viewer/}}, and the following software: CLASS \citep{Lesgourgues2011}, Jupyter Notebook \citep{Kluyver2016jupyter},
 Mathematica \citep{Mathematica}, Matplotlib \citep{Hunter2007}, NumPy \citep{VanderWalt2011}, Python \citep{Millman2011, Oliphant2007}, scikit-learn \citep{Pedregosa2012}
% \software{Galacticus \citep{Benson2012}
%IPython \citep{Perez2007},

\end{acknowledgments}

\appendix

\section{Appendix}\label{sec:appendix}

The thermal relic WDM transfer functions can be approximated by an analytical function with a dependence on $\mth$, as show in equation A8 of \cite{Bode2001}, and Eq. \ref{eq:wdm_transfer_f}. We use the more recent fit to the characteristic length scale factor $\alpha(\mth)$ obtained by \cite{Viel2005, Viel2012}, with $\mu=1.12$:
% these numbers are for Bode2001
% with $\mu=1.2$
% \begin{equation}
% \begin{split}
%   \alpha =& 0.048 \left(\frac{\Omega_{\textrm{thWDM}}}{0.4}\right)^{0.15}\left(\frac{h}{0.65}\right)^{1.3} \times \\
%   &\left(\frac{\textrm{keV}}{\mth}\right )^{1.15}\left(\frac{1.5}{\gth}\right)^{0.29}
% \end{split}
% \end{equation}

\begin{equation}\label{eq:alpha}
\begin{split}
   \alpha &= 0.049
   \left(\frac{\mth}{\textrm{keV}}\right )^{-1.11}
   \left(\frac{\Omega_{\textrm{thWDM}}}{0.25}\right)^{0.11}\left(\frac{h}{0.7}\right)^{1.22}\equiv\lfs\\
   %&\equiv\lfs
\end{split}
\end{equation}
in units of Mpc $h^{-1}$. By convention, the number of degrees of freedom is taken to be 1.5 for the warm particle, based on the equivalent contribution of a light neutrino species \cite{Bode2001}.
Following \citep{Schneider2012}, we assume that the characteristic length scale $\alpha$ can be related to an effective free-streaming length scale $\lfs$. 
%The free streaming mass scale defines the scale at which initial density perturbations are erased

The `half-mode' length scale $\lhm$ corresponds to the scale at which the thermal relic WDM transfer function (\ref{eq:wdm_transfer_f}) decreases to 1/2: 
\begin{equation}
    \lhm = 2 \pi \lfs (2^{\mu/5} -1 )^{-\frac{1}{2\mu}}.
\end{equation}
We define the corresponding `half-mode' mass scale and obtain a relation between $\mhm$ and $\mth$:
\begin{equation}
\begin{split}
   \mhm =& \frac{4\pi}{3}\bar{\rho} \left(\frac{\lhm}{2}\right)^3 =  \frac{4\pi^4}{3} \bar{\rho} \left(2^{\mu/5} -1 \right)^{-\frac{3}{2\mu}} {\lfs}^{~3}\\
   %=& \frac{4\pi^4}{3} \bar{\rho} \left(2^{\mu/5} -1 \right)^{-\frac{3}{2\mu}}   0.049^3
   %\left(\frac{\Omega_{\textrm{thWDM}}}{0.25}\right)^{0.33} \times \\
   %&\left(\frac{h}{0.7}\right)^{3.66} \left(\frac{\mth}{\textrm{keV}}\right )^{-3.33}\\
    =& A \left(\frac{\mth}{3.3\textrm{keV}}\right )^{-3.33},
\end{split}
\end{equation}
where $\bar{\rho}$ is the background density of the universe. The
$A$ constant thus depends on the assumed  $\bar{\rho}$. 

We consider two cases, where the background density of the universe can be taken as:
\begin{enumerate}[label=\Roman*.]
%\begin{enumerate}
	\item $\bar{\rho} = \Om \times \rho_{\textrm{critic}}$, where $\Om$ includes the contributions of both baryonic and dark matter (as done in \cite{Nadler2021}, or 
	\item $\bar{\rho} = \Odm \times \rho_{\textrm{critic}}$, where only the contributions of dark matter are included (as done in \cite{Gilman2020}).
\end{enumerate}

We thus obtain two equations between the `half-mode' mass and the thermal relic mass:

\begin{equation}
\begin{split}
       &\textrm{I}.~ \mhm= \AI \left(\frac{\mth}{3.3\textrm{keV}}\right )^{-3.33} M_{\odot}\\ 
       &\textrm{II}.~ \mhm= \AII \left(\frac{\mth}{3.3\textrm{keV}}\right )^{-3.33} M_{\odot}\\
\end{split}
\end{equation}

% \textbf{A more recent work by \cite{Viel2012}  obtained a formula that models the non-linear evolution of thermal relic transfer function as a function of redshift. This model can be taken into account in future work when analyzing the data obtained from individual lenses/objects, when marginalizing over their individual redshift posteriors. }
\clearpage
\nocite{*}
\bibliography{dm}% Produces the bibliography via BibTeX.


\end{document}